\documentstyle[12pt,twoside]{article}
\begin{document}
\begin{center}

{\Large\bf The clustering of dark energy\\[5PT]}
\medskip
 
{\bf F.G. Alvarenga\footnote{e-mail: flavio@cce.ufes.br},
J. C. Fabris\footnote{e-mail: fabris@cce.ufes.br}, S.V.B. Gon\c{c}alves\footnote{e-mail: sergio@cce.ufes.br} and
G. Tadaiesky\footnote{gtadaiesky@cce.ufes.br}} \medskip

Departamento de F\'{\i}sica, Universidade Federal do Esp\'{\i}rito Santo, 
29060-900, Vit\'oria, Esp\'{\i}rito Santo, Brazil \medskip\\
\medskip

\end{center}
 
\begin{abstract}

A cosmological model with pressurelles matter and a fluid of negative pressure
is studied. At perturbative level, fluctuations of both fluids are considered.
It is shown that at least at very large scales, the fluid of negative pressure,
which represents the dark energy content of the universe, clusters like the
dust fluid. Numerical integration reveals that this behaviour may also occur at scales
smaller than the Hubble radius.
\vspace{0.7cm}

PACS number(s): 95.35.+d, 98.65.Dx
\end{abstract}
 
The dynamics of clusters of galaxies suggests the presence of a large amount of
dark matter \cite{kent}. The mass-luminosity ratio for these systems, which may take values of the
order of $200$ in solar unities, implies that the fraction density of dark matter
is such that $\Omega_{dm} \sim 0.3$. On the other hand, cosmological data coming, for example,
from the position of the first accoustic peak in the spectrum of the anisotropy of the cosmic
microwave background radiation, suggests that the universe is almost spatially flat \cite{charles}.
Hence, about $70\%$ of the matter content of the universe remain a smooth unclustered matter
component since
it does not appear in the dynamics of clusters of galaxies. This smooth component is
generally called "dark energy".
\par
A fluid of negative pressure is generally evoked to represent the dark energy of the universe.
The cosmological constant is a particular case, characterized by $p = - \rho$.
When the negative pressure is dynamically generated by a self-interacting scalar
field, this component is popularly called "quintessence" \cite{steinhardt}.
The behaviour of fluctuations in fluids with negative pressure,
in the hydrodynamic representation, was investigate in
\cite{jerome}, where it has been showed that there are only decreasing modes in the
long wavelength approximation. This fact justifies the using of fluids of negative pressures
to represent the smooth component of the matter content of the universe.
\par
In this brief note we reanalyze the fate of density perturbations in fluids with negative pressure.
However, in order to keep the model closer to a realistic one, a two-fluid system is studied.
Notice that since $\Omega_{dm} \sim 0.3$ and $\Omega_{de} \sim 0.7$, the universe today has
not a component that dominates completly the background.
One of the components is a pressurelles fluid which is censed to represent dark matter;
the other is a fluid described by the barotropic equation of state $p = \alpha\rho$, with
$\alpha < 0$, representing the dark energy. In the present approach both fluids fluctuates,
what is the more general case from both the physical and mathematical points of view.
In many studies in the literature on this subject(see for example \cite{lokas}), only the dark matter component is
perturbed, since it is assumed that the other component remains unclustered, an assumption
justified by the behaviour of the one-fluid model for fluids with negative pressure.
\par
Here, it will be shown that, at least at very large scales, the fluid of negative pressure also
clusters. Using numerical integration, it is possible to verify that the clustering of
the negative pressure fluid occurs for scales smaller than the Hubble radius. If a smooth
component is necessary at scales up to Hubble radius, only the cosmological constant is
a serious candidate. Another possibility, as it will be discussed later, is the Chaplygin
gas, a very particular fluid with negative pressure.
\par
Let us consider the Einstein's equations in presence of two fluids which do not interact
between themselves, but only through geometry. One of the fluid has no pressure and the
other has the general equation of state $p = \alpha\rho$. Using the Friedmann-Robertson-Walker flat
metric, we obtain the equation governing the evolution of the universe:
\begin{equation}
\label{fe}
\biggr(\frac{\dot a}{a}\biggl)^2 = \biggr(\frac{\Omega_{m0}}{a^3} + \frac{\Omega_{q0}}{a^{3(1 + \alpha)}}\biggl) \quad ,
\end{equation}
where $\Omega_{m0}$ and $\Omega_{q0}$ represents the fraction density of each of these
components today with respect to the total density.
\par
From (\ref{fe}), we may write
\begin{equation}
\dot a = \biggr\{\frac{\Omega_{m0}}{a} + \frac{\Omega_{q0}}{a^{1 + 3\alpha}}\biggl\}^{1/2} \equiv F(a) \quad .
\end{equation}
In this expression, we have parametrize the time in terms of the Hubble time $H_0^{-1}$,
and used the fact that $\frac{8\pi GH_0^{-2}\rho_{T0}}{3} = 1$, $\rho_{T0}$ being the total density of
the universe today.
Using standard techniques \cite{weinberg,peebles,turner}, the Einstein's equations are perturbed and only the scalar sector
is considered since it is directly connected with density perturbations.
In the synchronous coordinate condition, the perturbed equations reduce to
\begin{eqnarray}
\label{pe1}
\Delta_m'' + \biggr(\frac{F'}{F} + \frac{2}{a}\biggl)\Delta_m' &=& \frac{3}{2a^2F^2}\biggr(
\frac{\Omega_{m0}}{a}\Delta_m + (1 + 3\alpha)\frac{\Omega_{q0}}{a^{1 + 3\alpha}}\Delta_q\biggl) \quad ,
\\
\label{pe2}
\Delta_q' + (1 + \alpha)\biggr(\frac{\Theta}{F} - \Delta_m'\biggl) &=& 0 \quad , \\
\label{pe3}
(1 + \alpha)\biggr(\Theta' + (2 - 3\alpha)\frac{\Theta}{a}\biggl) &=& \alpha \frac{k^2}{a^2F}\Delta_q \quad .
\end{eqnarray}
In these equations $\Delta_m$ and $\Delta_q$ are the density contrast for ordinary matter and the dark energy respectivelly, $\Delta_i = \frac{\delta\rho_i}{\rho_i}$. The divergence of the perturbation in the four-velocity of the dark energy fluid
was denoted by $\Theta$, and $k$ is the wavenumber of the perturbation. The primes mean
derivative with respect to the scale factor $a$. In deriving
(\ref{pe1},\ref{pe2},\ref{pe3}) it has been used the fact that $\Delta_m = \frac{h}{2}$, where
$h$ is the trace of metric perturbations, a relation coming from the equation of conservation
for the pressurelles fluid.
\par
Strictly speaking, if one has in mind a kind of quintessence scenario, a self-interacting
scalar field should be considered instead of a hydrodynamical fluid with negative pressure.
But, as it was discussed in \cite{sergio}, for long wavelengths the hydrodynamical approach
leads to results similar to the scalar field representation.
\par
In the long wavelenght limit, one can fix $k = 0$. Hence, the above equations
reduce themselves to
\begin{eqnarray}
\label{fe1}
\Delta_m'' + \biggr(\frac{F'}{F} + \frac{2}{a}\biggl)\Delta_m' &=& \frac{3}{2a^2F^2}\biggr[
\frac{\Omega_{m0}}{a} + (1 + 3\alpha)\frac{(1 + \alpha)}{a^{1 + 3\alpha}}\biggl]\Delta_m  \quad , \\
\label{fe2}
\Delta_q &=& (1 + \alpha)\Delta_m \quad ,
\end{eqnarray}
where decreasing inhomogeneous term, coming from the decoupled variable $\Theta$, were discarded.
From, (\ref{fe2}) it can be verified that density contrast for the dark energy component follows
the same behaviour of the pressurelless fluid.
\par
Equation (\ref{fe1}) may be reduced to the hypergeometric equation
\begin{equation}
y(1 - y)\ddot\lambda - \biggr[\frac{5 - 6\alpha}{6\alpha} - \frac{5 - 9\alpha}{6\alpha}y\biggl]\dot\lambda + \frac{5 + 3\alpha}{6\alpha}\lambda = 0 \quad ,
\end{equation}
where $\Delta_m = a\lambda$, $y = - Aa^{-3\alpha}$ and $A = \Omega_{q0}/\Omega_{m0}$.
There is one growing solution. In terms of the density contrast, it reads
\begin{equation}
\Delta_m \propto a\;_2F_1(-\frac{5+3\alpha}{6\alpha},1,-\frac{5-6\alpha}{6\alpha},
-\frac{1 - \Omega_{m0}}{\Omega_{m0}}a^{-3\alpha}) \quad .
\end{equation}
In the long wavelength limit the density contrast for the pressurelless fluid and for the
quintessence fluid exhibit a growing mode. Hence, both component clusters.
\par
This result has been obtained in the long wavelength limit. For the analysis of
structure formation in such simplified model, finite wavelengths must be considered,
mainly those that are smaller or of the order of the Hubble radius. It seems not possible
to obtain closed expressions
for any value of $k$. Hence, we must turn to numerical integration. In terms of the cosmic
time $t$ the two-fluid perturbed system may be reduced to
\begin{eqnarray}
\ddot\Delta_m + 2\frac{\dot a}{a}\dot\Delta_m - \frac{3}{2}\biggr(\frac{\dot a}{a}\biggl)^2\frac{1}{F^2}\frac{\Omega_{m0}}{a}\Delta_m &=&
\frac{3}{2}(1 + 3\alpha)\biggr(\frac{\dot a}{a}\biggl)^2\frac{1}{F^2}\frac{\Omega_{q0}}{a^{1+3\alpha}}\Delta_q \quad ,
\\
\ddot\Delta_m + (2 - 3\alpha)\frac{\dot a}{a}\dot\Delta_m &=& \frac{1}{1 + \alpha}\biggr\{\ddot\Delta_q + (2 - 3\alpha)\frac{\dot a}{a}\dot\Delta_q + \alpha\biggr(\frac{k}{a}\biggl)^2\Delta_q\biggl\} \quad .
\end{eqnarray}
This system of differential equations may be analyzed, for example, in two asymptotics regimes:
when pressurelless matter dominates, leading to $a \propto t^\frac{2}{3}$; when the fluid of negative
pressure dominates, leading to $a \propto t^\frac{2}{3(1 + \alpha)}$. The numerical integration, with $k \neq 0$, indicates that, in the matter dominate period, which can be considered as
representing the evolution of the universe until $z \sim 1$, both fluids agglomerate.
However, for large values of $k$, i.e., small wavelength, or in the case where
the exotic fluid dominates completly the background, instabilities appears due to
hydrodynamical representation of the fluid of negative pressure employed here.
For this case, the
scalar field representation must be employed. In this high frequency regime, it must
be expected that the pressure of the
exotic fluid becomes positive \cite{grisha}. Hence, the question of the clustering of the exotic fluid
becomes more delicate. Moreover, at this limit, non-linear effects must be taken into
account. Studies performed with quintessential models using spherical symmetries suggest
that clustering may occur in this small wavelength limit \cite{wetterich}.
\par
A complete smooth dark energy is obtained for the cosmological case. In fact, with
$\alpha = - 1$, $\Delta_q = 0$, as it should be expected. Another example of dark
energy that remains essentially unclustered is the Chaplygin gas \cite{patriciabis}. The Chaplygin gas \cite{kamen}
is characterized by an equation of state $ p = - \frac{B}{\rho}$, where $B$ is
a constant. It has an interesting connection with the Nambu-Goto action for a
d-brane in a $d+2$ dimensional space-time. Due to its particular equation of state,
the Chaplygin gas exhibits an initial behaviour similar to a dust fluid for small
values of the scale factor, approaching later a behaviour close to a cosmological
constant. Hence, it initially clusters, becoming later a smooth component.
This behaviour is verified even in presence of ordinary matter \cite{patricia}.
\par
The main goal of the present work is to call attention that, in two-fluid model,
fluids with negative pressure exhibit a clustering behaviour, perhaps even in scales
below the Hubble radius. The clustering of dark energy has already been studied in
some other situations, like a quintessence model with a non-minimal coupling
between the scalar-field and the geometry, where the fluctuations in the geometry
lead to a dragging effect on the scalar field representing quintessence, due
to the non-minimal coupling \cite{perrotta}. But, using a very simple model, with
two barotropic fluids, it has been shown here that the clustering may also occurs
in other situations, provided the fluctuations in all components are taken into
account, what is the more natural scenario.
This, evidently, has important consequences for the structure
formation problem, and even for the anisotropy of the cosmic microwave background radiation.
\vspace{0.5cm}

{\bf Acknowledgement:} We thank CNPq (Brazil) for partial financial support.

\end{document}